\documentstyle[11pt,newpasp,twoside,psfig,epsf]{article}
\begin{document}

\title{VLT Observations of Two Unusual BAL Quasars}
\author{Patrick B. Hall}
\affil{Princeton University Observatory, Princeton, NJ 08544, USA%\\
%and Departamento de Astronom\'{\i}a y Astrof\'{\i}sica, Facultad
%de F\'{\i}sica, Pontificia Universidad Cat\'{o}lica de Chile,
%Casilla 306, Santiago 22, Chile
}
\author{Damien Hutsem\'ekers}
\affil{Research Associate FNRS, University of Li\`ege, Belgium}

\setcounter{page}{111}
% this page number will be filled later by the editors....
\index{Hall, P. B.}
\index{Hutsem\'ekers, D.}

\begin{abstract}
Among the unusual broad absorption line quasars being found by the %The
Sloan Digital Sky Survey (SDSS) are %is finding numerous
objects %broad absorption line quasars 
with much stronger absorption in
%Al\,{\sc iii} than Mg\,{\sc ii}, or much stronger 
Fe\,{\sc iii} than Fe\,{\sc ii}.  
These unusual line ratios require a high density in 
the outflow ($n_H \geq 3 \times 10^{10}$ cm$^{-3}$). 
They should also appear for only a limited range of
outflow column densities, which explains their rarity.
Previously we suggested that the Fe\,{\sc iii} line ratios were also
affected by a resonance; we now believe %Here we show that 
this is %instead %probably
an artifact of structure in the underlying %continuum.
Fe\,{\sc ii}+Fe\,{\sc iii} pseudocontinuum.
The SDSS is also discovering objects with absorption in rarely seen
transitions such as He\,{\sc i}.
VLT+UVES high-resolution spectra %spectroscopy
of one such object, %of them,
the mini-BAL %mini-LoBAL %low-ionization mini-BAL
quasar SDSS 1453+0029, %.  This object
show that it has two %rare 
He\,{\sc i} absorption systems %separated by only 350 km/s,
%but with different properties in each system.
with considerably different properties %in each system.
separated by only 350 km s$^{-1}$. %km/s.

\end{abstract}

\section{A Luminous Fe\,{\sc iii}-dominant BAL Quasar}
\noindent{Broad Absorption Line (BAL) quasars show absorption from gas with blueshifted
outflow velocities of typically 0.1$c$.  About 10\% %13\%
of quasars exhibit BAL troughs, %(Reichard et al. 2002).  This is
usually attributed either to an orientation effect such that all
quasars have BAL outflows covering $\sim$10\% %3\%
of the sky, or a phase of $\sim$100\% covering lasting $\sim$10\% %3\%
of the typical quasar lifetime.  Understanding BAL outflows is 
necessary %worthwhile
since the BAL outflow mass loss rates seem comparable to the accretion
rates required to power quasars. %($\sim1$\, $M_{\odot}$\,yr$^{-1}$).
%Either way, the BAL outflow mass loss rates seem comparable to the accretion
%rates required to power quasars ($\sim1$\, $M_{\odot}$\,yr$^{-1}$).
%Therefore an understanding
%of BAL outflows is required for an understanding of quasars as a whole.
%}
%
Unusual BAL quasars delineate %aid in this undertaking by delineating 
the full range of physical conditions and parameter space spanned
by BAL outflows.  The 
Sloan Digital Sky Survey (York et al. 2000; www.sdss.org) %(SDSS)
has %confirmed the existence of 
discovered populations of unusual
low-ionization broad absorption line quasars (`LoBALs'), %; Hall et al. 2002), 
many of them `FeLoBALs' with absorption from excited 
Fe\,{\sc ii} or Fe\,{\sc iii} (Hall et al. 2002).
}

%\bigskip
\begin{figure}[t]	%10cm 8.61cm
\centerline{ \psfig{figure=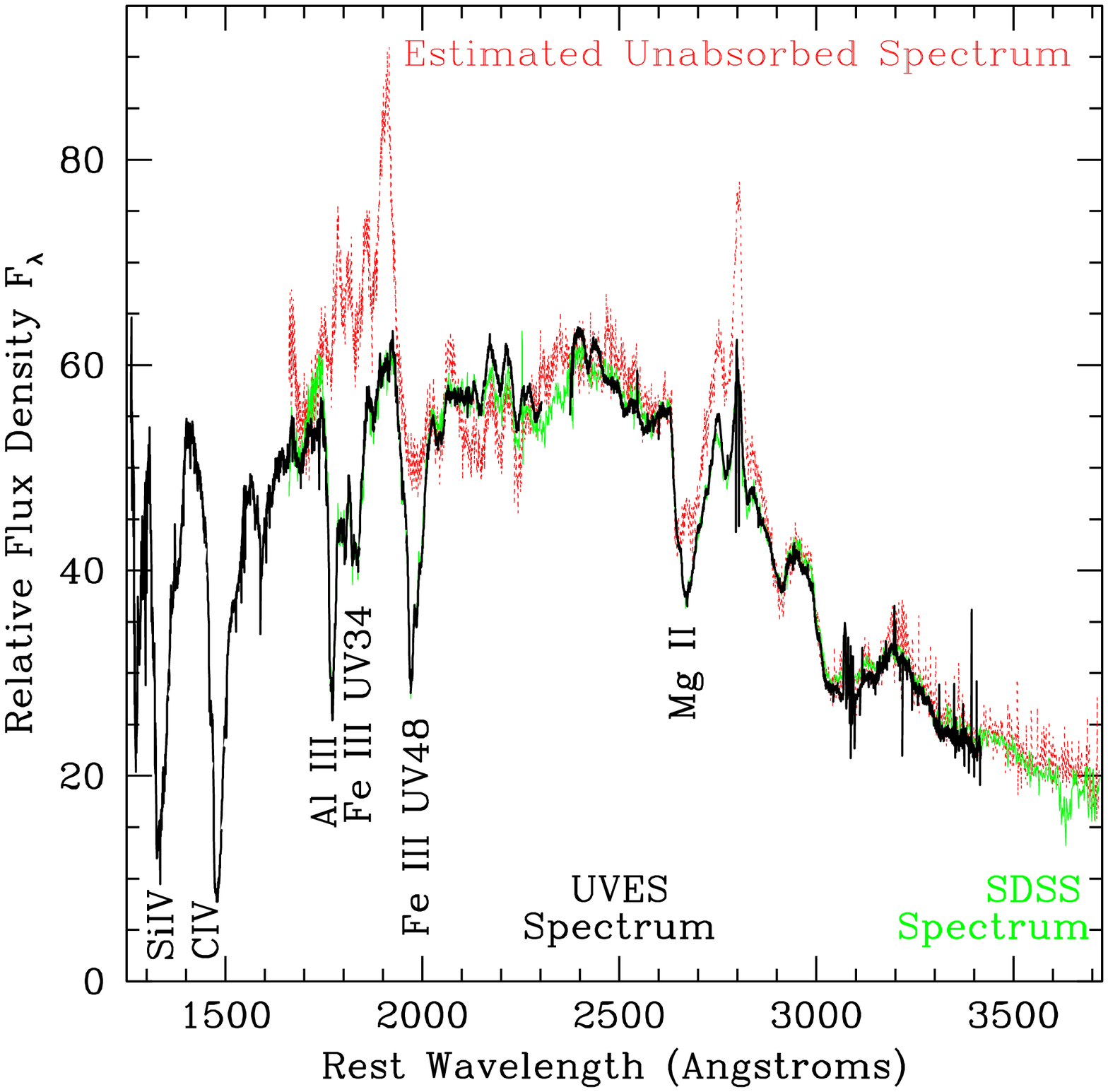,width=8.61cm,angle=0} %} %\centerline{
\psfig{figure=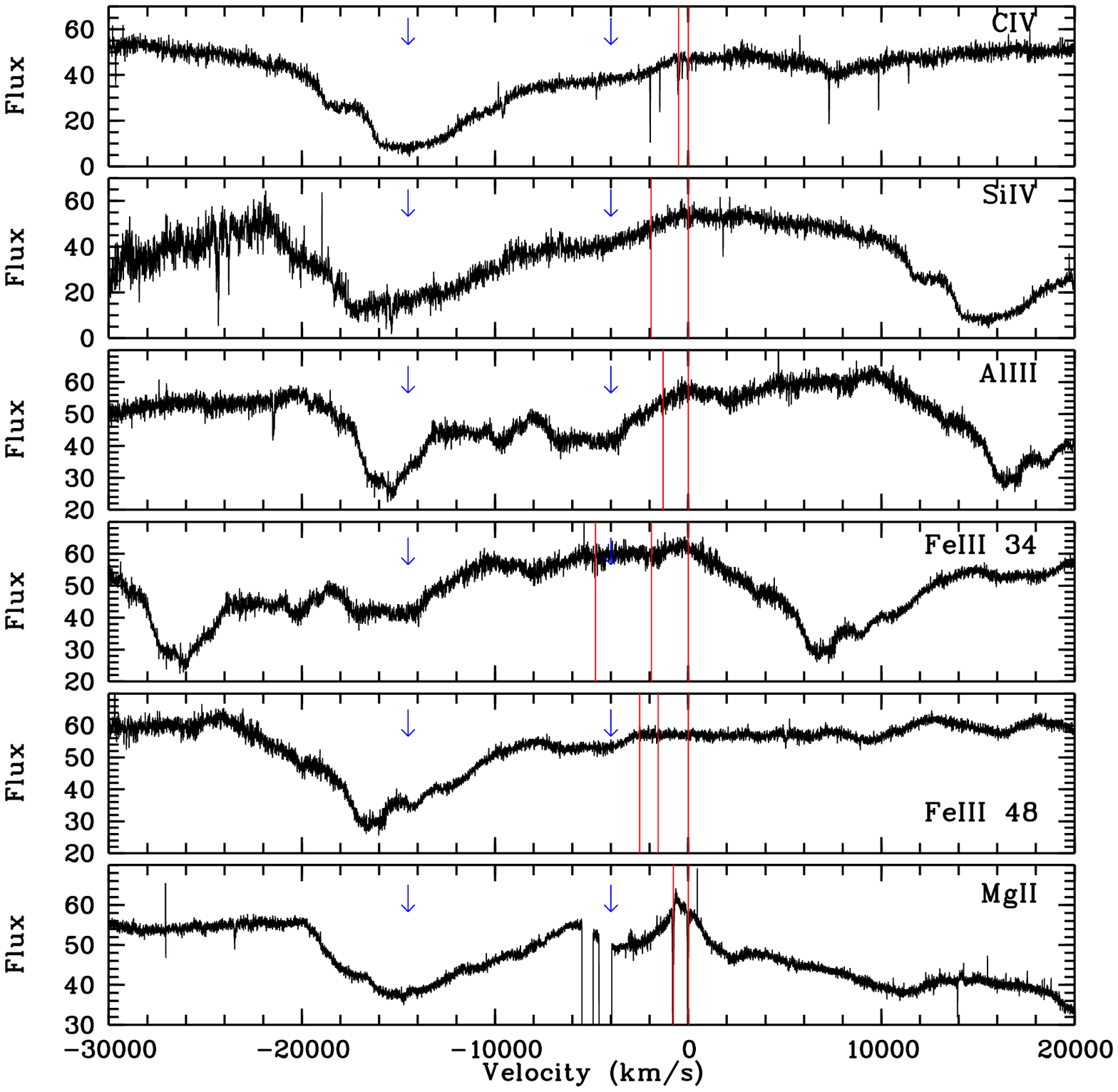,width=8.61cm,angle=0} }
\caption{UVES spectra of the Fe\,{\sc iii}-dominant BAL quasar SDSS 2215-0045,
a) compared with a model for its %underlying
unabsorbed spectrum, and
b) showing various absorption troughs on the same velocity scale.}
%(heavy black line), %and an estimate of its %intrinsic,
%unabsorbed spectrum (light red line).}
\label{fig11}
\end{figure}

%\section{A Luminous Fe\,{\sc iii}-dominant BAL Quasar}
One population of unusual BAL quasars has unprecedented line ratios which are
the reverse of those seen in %all
previously known objects; namely, %consists of
LoBALs with stronger absorption in Al\,{\sc iii} than Mg\,{\sc ii}
and FeLoBALs with %at best very
weak or no Fe\,{\sc ii} absorption but with %absorption in
Fe\,{\sc iii} even stronger than Mg\,{\sc ii} (Fig. 1a).
The CLOUDY modeling of de Kool et al. 2002 (his Figs. 6-7)
shows that these ratios can be produced in %are due to 
a high density outflow
(log $n_H$$\geq$10.5 cm$^{-3}$ for ionization parameter $U\simeq-2$)
%(e.g., log $N_H$=22.3$\pm$0.1 for log $n_H$=11),
with column density $N_H$ in a narrow range ($\sim$0.2 dex) such that
a partially ionized zone of Fe\,{\sc ii} and Mg\,{\sc ii} %emitting zone 
--- typically present in AGN broad emission line regions ---
does not exist in the BAL outflow along our line of sight,
but a Fe\,{\sc iii}+Al\,{\sc iii} zone does.  
%but a zone of Fe\,{\sc iii} and Al\,{\sc iii} %emitting zone 
%does.
%(de Kool {\em et\,al.} 2002, ApJ, 570, 514, Fig. 6).
%	These BAL outflows can thus be thought of as {\em density-limited}.
The high densities and
narrow range in $N_H$ required to reproduce the observed
line ratios explains why these objects are so rare.
The SDSS has recently found %next SDSS data release includes 
a quasar with a narrow
Fe\,{\sc iii}-dominant absorption system for which high-resolution spectra
%spectroscopy
could test %confirm or refute 
this %proposed 
explanation. %--- %the density-limited hypothesis.
Unfortunately, %VLT+UVES data on
the most luminous Fe\,{\sc iii}-dominant BAL quasar known,
SDSS 2215-0045, cannot %be used to 
test this model since a %high-resolution
VLT+UVES spectrum shows no narrow features. %(Fig. 1b).
%
%%We obtained 
%VLT+UVES data on the %most
%luminous Fe\,{\sc iii}-dominant BAL quasar %known, 
%SDSS 2215-0045 in an unsuccessful search for %, to search for %shows no
%narrow features at high resolution (Fig. 1b).
%
However, the UVES spectrum does reveal
that C\,{\sc iv} and Si\,{\sc iv} emission are as weak as Al\,{\sc iii}.
This might be due in part to a weak %an
absorption trough at $\sim$4000\,km\,s$^{-1}$
in addition to the main trough at $\sim$15000\,km\,s$^{-1}$
(Fig. 1b, blue arrows).

%In SDSS 2215-0045,
The Fe\,{\sc iii} UV48 $\lambda$2080\AA\ absorption 
in SDSS 2215-0045
appears stronger than the UV34 $\lambda$1910\AA\ absorption,
which is unphysical.  In Hall et al. 2002 this was attributed to a previously 
unknown resonance populating the lower term of UV48, but we now believe it is 
due to a complicated underlying continuum.  As seen in Fig. 1a,
the unabsorbed continuum of 2215-0045 (thin dotted red line)
can be estimated as a power law plus Fe\,{\sc ii}+Fe\,{\sc iii}
emission (using the extreme Fe %Fe\,{\sc ii}
emitter SDSS 0923+5745 as a template),
all reddened by %$E$$(${\sc b}-{\sc v}$)$=0.06. 
$E$$($$B$-$V$$)$=0.06. %(Fig. 1, light red line).
Due to the strong emission at %rest 
1800\,\AA\ and the lack of emission at 2000\,\AA, the observed spectrum 
(heavy line) %then 
appears consistent with saturated Fe\,{\sc iii}
UV48 and UV34 absorption with 35\% partial covering.
The lower-ionization Mg\,{\sc ii} absorption has smaller partial covering,
which is typical. 
The Fe-emission template %Fe\,{\sc ii}-template model 
is not a perfect fit %,
(e.g., 2215-0045 has weaker Mg\,{\sc ii} and C\,{\sc iii}] emission
than the template),
but it shows the spectrum can be fit without invoking a new
Fe\,{\sc iii} resonance.

%-------------------------------------------------------------------------------

%VLT+UVES high-resolution spectra %spectroscopy 
%of the mini-LoBAL %low-ionization mini-BAL 
%quasar SDSS 1453+0029 %.  This object 
%show that it has two rare He\,{\sc i}
%absorption systems separated by only 350 km/s, 
%but with different properties in each system.

%\bigskip
\begin{figure}[b]
\centerline{
\psfig{figure=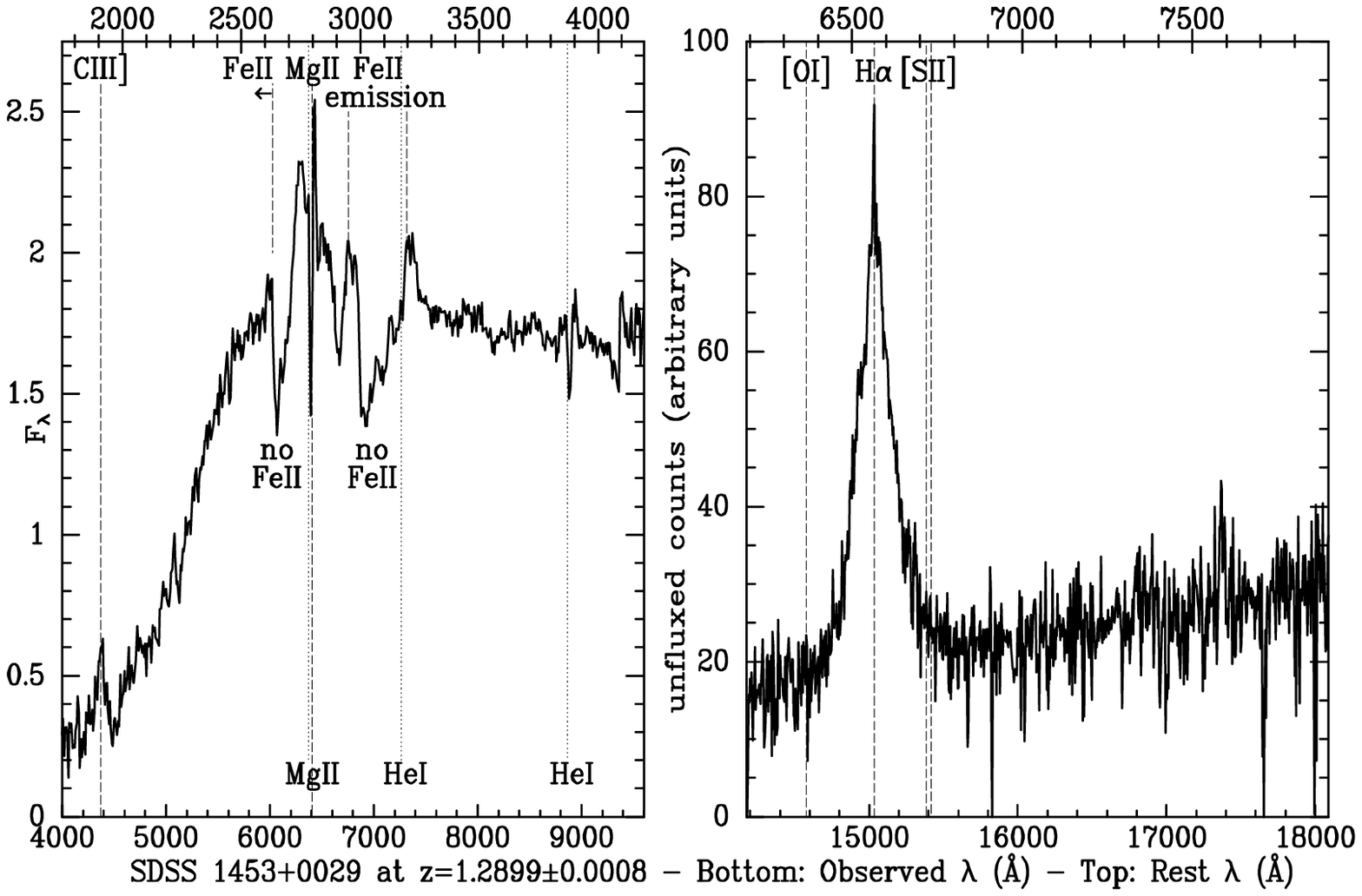,width=14.13cm,angle=0}
}
%\caption{Optical \& ISAAC\,NIR spectra of the mini-LoBAL\,1453+0029.}
\caption{Optical \& VLT+ISAAC NIR spectra of the mini-LoBAL 1453+0029. %}
The narrow H$\alpha$ $z$ agrees with the UVES absorption $z$.}
%\caption{Optical spectrum of the mini-LoBAL 1453+0029.}
%(heavy black line),
%and an estimate of its %intrinsic,
%unabsorbed spectrum (light red line).}
\label{fig21}
\end{figure}

%\bigskip
\begin{figure}[t]
\centerline{
\psfig{figure=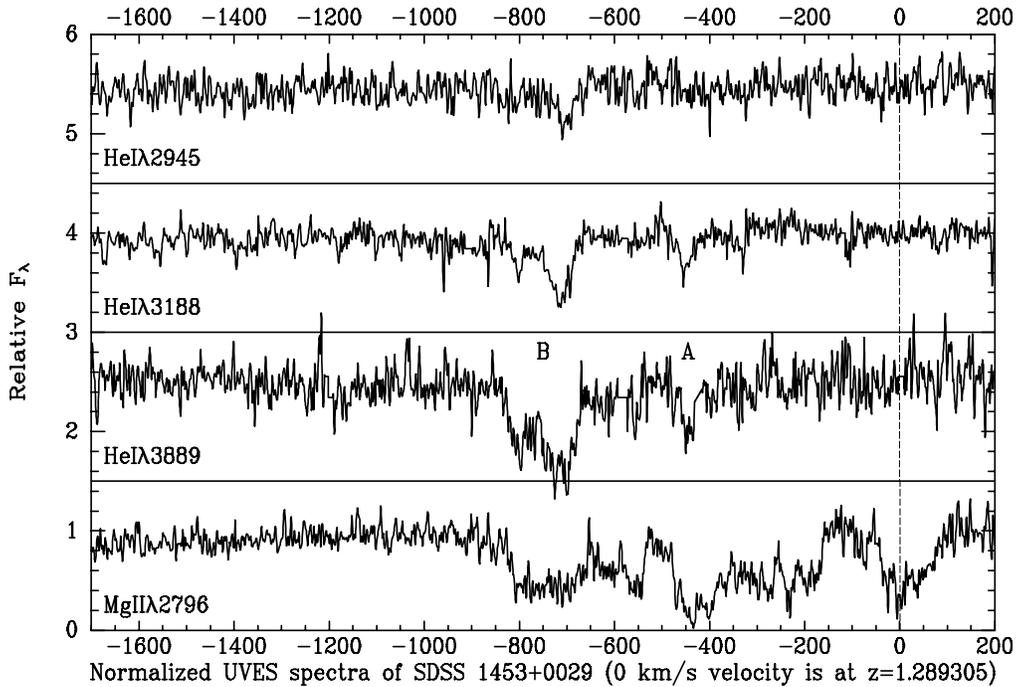,width=14.13cm,angle=0}
}
\caption{Important transitions in the UVES spectrum of %SDSS
1453+0029.}
%(heavy black line),
%and an estimate of its %intrinsic,
%unabsorbed spectrum (light red line).}
\label{fig22}
\end{figure}

%\section{VLT+UVES Spectra of a Reddened Mini-LoBAL Quasar}
\section{A Reddened Mini-LoBAL Quasar}
%\noindent{Broad Absorption Line (BAL) quasars show absorption troughs
%typically 10,000 km/s wide from outflowing ionized gas.
%%from gas with blueshifted outflow velocities typically spanning 10,000 km/s. 
\noindent{Traditionally, objects with troughs $<$2000 km/s wide, or with 
absorption solely within 3000 km/s of the quasar redshift, have not been
classified as BAL quasars.  
However, numerous studies have shown that both `mini-BALs' and 
many narrow associated systems also arise in the central regions of AGN.
Thus in Hall et al. (2002) %; ApJS 141, 267)
we suggested an alternative to the traditional balnicity
index which does not discriminate against narrow intrinsic absorbers.
The physics of AGN outflows are more easily studied %easier to study 
with such objects, since in BALs most doublet transitions are blended and 
cannot be used to infer the physical conditions or 
other~parameters~of~the~outflow.~For~example,~studies~of~numerous~intrinsic~AGN
outflows~have~shown~that~the~troughs~are~usually~saturated.~In~such~cases,~the
trough depth %of a trough
in high-resolution spectra yields %the covering factor of that ion 
that ion's covering factor
as a function of velocity.
Then, knowing the relative strengths of the various
transitions observed, 
interesting conclusions can be drawn by examining the spectra.
%transitions~observed,~interesting~conclusions~can~be~drawn~just~by~looking~at~the~spectra.
}

%%Whether there are intrinsic differences between narrow and broad AGN outflows
%%is something the Sloan Digital Sky Survey (SDSS) should be able to address.
%In Hall et al. (2002; ApJS 141, 267)
%we suggested an alternative to the traditional balnicity
%index which does not discriminate against narrow intrinsic absorbers.
%One interesting example of such a system occurs in the reddened,
%extreme-Fe\,{\sc ii}-emitting, low-ionization mini-BAL quasar SDSS 1453+0029.

An excellent example is %UVES spectra of 
the reddened, extreme-Fe\,{\sc ii}-emitting, mini-LoBAL 
quasar SDSS 1453+0029, %illustrate this (Fig. 2).
which has $z$=1.2899$\pm$0.0008 from %a VLT+ISAAC 
an ISAAC 
near-IR spectrum (Fig. 2).  Broadly speaking, this object %it
has~two He\,{\sc i} absorption systems within a more extensive
(in velocity space) set of Mg\,{\sc ii} absorption systems.
%
%This He\,{\sc i} absorption consists of three lines.  
%The lower state giving rise to this He\,{\sc i} absorption 
%is an excited %a triplet
This He\,{\sc i} absorption arises in an excited
state populated by recombination
(He\,{\sc ii} + e$^{-}$ $\rightarrow$ He\,{\sc i} + $\gamma$).
Thus this He\,{\sc i} absorption traces the column density of He\,{\sc ii},  
%Using the cosmic abundance of He then gives 
which yields a lower limit on $N_H$ %the H\,{\sc ii} column density
since hydrogen is completely ionized in He\,{\sc ii} regions.
Component A in SDSS 1453+0029
appears to have equally strong
He\,I$\lambda$3188 and He\,I$\lambda$3889 absorption (Fig. 3).
Since the latter transition is intrinsically much stronger,
this means that He\,I$\lambda$3889 is saturated.  
The covering factor of the He\,{\sc i} ($\sim$50\%) is 
less than that of Mg\,{\sc ii} ($\sim$85\%) by inspection.
%
%**but if He\,{\sc i} is unsaturated, the covering factor could be equal? even greater?
%
Component B has He\,{\sc i} line ratios consistent with He\,I$\lambda$3889 being
at most only slightly saturated, but its velocity-dependent covering factor
reaches a maximum $\sim$90\% while 
that of Mg\,{\sc ii} is only $\sim$60\%.
%Mg\,{\sc ii} has a covering factor of $\sim$60\%.
%
%Thus %in
%%SDSS 1453+0029 has %there are
%%two absorption systems %separated by 350\,km\,s$^{-1}$
%%wherein the He\,{\sc i} has very different parameters from the Mg\,{\sc ii}.
%the He\,{\sc i} behaves very differently %has very different parameters from
%from Mg\,{\sc ii} in %between 
%the two different absorption systems in SDSS 1453+0029.
%%where the He\,{\sc i} has very different properties relative to the Mg\,{\sc ii}.
%
%The two He\,{\sc i} absorption systems in SDSS 1453+0029 show that
%low- and high-ionization absorption can behave very differently
%even over small $\Delta v$.%velocity ranges.
%
%Despite occurring at similar velocities, and thus very likely %probably
%in physical proximity,
%%indicating linkage of at least some of the gas,
Thus, He\,{\sc i} and Mg\,{\sc ii} absorption can behave very differently,
even over small %$\Delta v$.%
velocity ranges, and even when
the two components are very likely in physical proximity
as indicated by their nearly identical velocities.
%such absorption occurs at nearly identical velocities which indicate 
%that the two components are very likely in physical proximity.
%
%%This is in spite of their similar velocities, which indicates that the He\,{\sc ii}
%%gas is in close physical proximity to at least some of the Mg\,{\sc ii} gas.

%-------------------------------------------------------------------------------

%\section{Conclusions} 

\acknowledgements
Based on data from ESO programs 267.A-5698 %program 267.A-5698.
and 67.A-580.
%
%The SDSS Web site is http://www.sdss.org/.  
Funding for %the creation and distribution of 
the SDSS Archive has been provided by
the Alfred P. Sloan Foundation, the Participating Institutions,
NASA, %the National Aeronautics and Space Administration, 
the NSF, %the National Science Foundation, 
the U.S. Department of Energy, 
the Japanese Monbukagakusho, and the Max Planck Society. 
%The SDSS Web site is http://www.sdss.org/.  
The SDSS is managed by the Astrophysical Research Consortium, % (ARC) 
for the Participating Institutions: %. The Participating Institutions are 
The University of Chicago, Fermilab, the Institute for Advanced Study,
the Japan Participation Group, The Johns Hopkins University, Los Alamos
National Laboratory, the Max-Planck-Institute for Astronomy, % (MPIA),
the Max-Planck-Institute for Astrophysics, % (MPA),
New Mexico State University, University of Pittsburg, Princeton University,
the USNO %United States Naval Observatory, 
and the University of Washington.

\end{document}